\documentclass[aps,prb,twocolumn,superscriptaddress,nofootinbib]{revtex4-2}
\usepackage{amssymb,graphicx,color}
\usepackage[intlimits]{amsmath}
\usepackage[english]{babel}
\usepackage[colorlinks]{hyperref}
\hypersetup{
	colorlinks=blue,
	linkcolor=blue,
	filecolor=blue,
	urlcolor=blue,
	citecolor=blue
}
\usepackage[normalem]{ulem}
\usepackage{comment}
\usepackage{ragged2e}
\usepackage{enumerate}
\usepackage{subfigure}
\usepackage{float}
\usepackage{ulem}
\usepackage{dsfont}

\newcommand{\ket}[1]{|\,{#1}\,\rangle}

\newcommand{\bref}[1]{(\ref{#1})}
\newcommand{\fref}[1]{Fig.~\ref{#1}}

\newcommand{\eref}[1]{Eq.~(\ref{#1})}

\newcommand{\sref}[1]{section~\ref{#1}}

\newcommand{\cref}[1]{chapter~\ref{#1}}

\newcommand{\Cref}[1]{Chapter~\ref{#1}}

\usepackage{ulem}[normalem]   
\begin{document}

\title{The disordered Dicke model}
	
\author{Pragna Das}
\affiliation{Indian Institute of Science Education and Research Bhopal 462066 India}
\author{Sebastian W\"{u}ster}
\affiliation{Indian Institute of Science Education and Research Bhopal 462066 India}
\author{Auditya Sharma}
\affiliation{Indian Institute of Science Education and Research Bhopal 462066 India}
	
\begin{abstract}
  We introduce and study the disordered Dicke model in which the
  spin-boson couplings are drawn from a random distribution with some
  finite width. Regarding the quantum phase transition we show that
  when the standard deviation $\sigma$ of the coupling strength
  gradually increases, the critical value of the mean coupling
  strength $\mu$ gradually decreases and after a certain $\sigma$
  there is no quantum phase transition at all; the system always lies
  in the super-radiant phase. We derive an approximate expression for
  the quantum phase transition in the presence of disorder in terms of
  $\mu$ and $\sigma$, which we numerically verify. Studying the
  thermal phase transition in the disordered Dicke model, we obtain an
  analytical expression for the critical temperature in terms of the
  mean and standard deviation of the coupling strength.  We observe
  that even when the mean of the coupling strength is zero, there is a
  finite temperature transition if the standard deviation of the
  coupling is sufficiently high. Disordered couplings in the Dicke
  model will exist in quantum dot superlattices, and we also sketch
  how they can be engineered and controlled with ultracold atoms or
  molecules in a cavity.
\end{abstract}

\maketitle 
		 
\section{Introduction}
\begin{figure}[t]
  \includegraphics[width=0.45\textwidth]{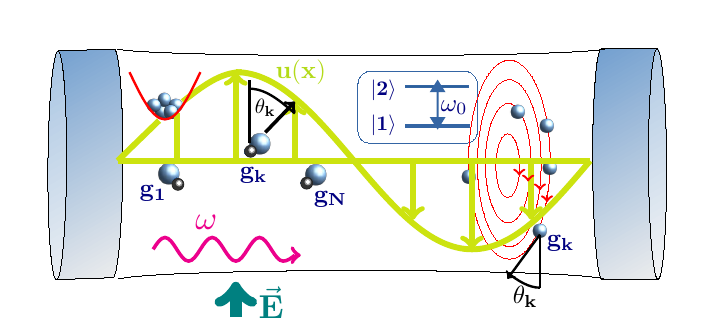}\label{fig:boson}
  \caption{ Schematic of the disordered Dicke model where $N$ two 
  levels atoms are coupled to a single mode bosonic field with different 
  spin-boson coupling strengths $g_k$, and two possible realisations.
The frequency of the bosonic mode is $\omega$ and the gap between two 
levels $|1\rangle$ and $|2\rangle$ of each atom is $\omega_0$. (left) 
These could be ultra-cold molecules whose fixed transition dipoles are 
randomly oriented wrt.~the cavity field direction (green), 
or (right) 
atoms under the influence of an additional external field that breaks 
their symmetry, such as the magnetic field around a wire (red). 
}
  \label{fig:schematic}
\end{figure}
The Dicke model~\cite{ dicke1954coherence}, which describes the
interaction between light and matter, is of fundamental importance
within the field of quantum optics. It exhibits a variety of
interesting phase transitions covering quantum phase
transitions~\cite{emary2003quantum, emary2003chaos,
  lambert2004entanglement, baumann2010dicke, kirton2019introduction}
(QPT), excited-state quantum phase transitions~\cite{perez2017thermal,
  das2022revisiting} (ESQPT) and thermal phase
transitions~\cite{wang1973phase, perez2017thermal,
  zhu2019entanglement, cejnar2021excited, das2022revisiting} (TPT).
The QPT 
takes place in the thermodynamic limit of infinite atom number,
$N\to\infty$, where the system goes from the normal phase (NP) to 
the super-radiant phase (SP)~\cite{ gross1982superradiance} at some
critical coupling strength $g_c$~\cite{emary2003chaos} between spins
and bosons. If temperature is introduced to the system, for $g>g_c$,
there is a critical temperature $T_c$, above which the system returns 
to the NP from the SP, whereas for $g<g_c$ the system lies in the
NP for all temperatures~\cite{carmichael1973higher, wang1973phase,
  duncan1974effect, quan2009quantum, Bastarrachea-Magnani_2017,
  perez2017thermal, das2022revisiting}. 
		  
Here, we generalize the standard Dicke model towards disorder in the
coupling strength $g$, for which we propose several practical
realisations.  While the role of disorder in the more general
spin-boson model has been considered both in
theoretical~\cite{temnov2005superradiance,
  marchetti2006thermodynamics, marchetti2007absorption,
  diniz2011strongly, tsyplyatyev2009dynamics, dhar2018variational} and
experimental~\cite{ krimer2014non, krimer2016sustained,
  zhang2018dicke} studies, the exploration of disorder-induced
phenomena within this context is still at a nascent stage. We focus on
those here, with the aid of tools from quantum information theory such
as mutual information~\cite{vedral2003classical,
  divincenzo2004locking, adesso2010quantum} between two spins as a
function of temperature, whose usefulness has been demonstrated for
clean Dicke models earlier~\cite{das2022revisiting, das2023phase}.

In the usual clean Dicke model, it is well known that the
QPT~\cite{emary2003quantum, emary2003chaos, lambert2004entanglement,
  baumann2010dicke, kirton2019introduction} occurs at some critical
light matter coupling strength. We find that for the disordered Dicke
model both the mean and the standard deviation of the random coupling
distribution play a crucial role in the QPT. If either one of them or
both are high, then the ground state exhibits super-radiant behaviour.
To show this, we numerically calculate the ground state energy and
average boson number as a function of the coupling distribution for
the disordered Dicke model. We verify our numerics with the aid of
available analytical results~\cite{emary2003chaos} for the ground
state energy and the average boson number across the QPT in the usual
Dicke model. By carrying out a disorder-averaging of their results, we
obtain an approximate expression for the critical line of the QPT in
the disordered model. The behaviour of the observables (ground state
energy and average boson number) around the critical line that we
obtained using a Taylor series expansion shows broad agreement with
our numerics. Moreover, we show how a symmetry of the Hamiltonian can
be exploited along with a heuristic argument to obtain the line of
quantum criticality in a more accurate manner.

To understand the thermal phase transition, we follow methods for
which the basis was laid in Ref.~\cite{ wang1973phase, hioe1973phase},
and calculate the partition function of the disordered Dicke model to
obtain the critical temperature in terms of the disorder coupling
strength. Numerically we calculate the mutual information between two
spins for the disordered Dicke model by a method similar to our
earlier work~\cite{das2022revisiting} for the usual Dicke model. When
the width of the disorder is sufficiently high, there is a finite
temperature transition from the SP to the NP even if the mean of the
coupling strength is zero.  We can predict the critical temperature of
this transition analytically, signatures for which are also seen in
the mutual information found numerically.

Earlier studies of disorder in the Dicke model considered the
multi-mode case~\cite{rotondo2015dicke}.  In contrast, we sketch
several possible realisations of disorder in the \emph{single mode}
Dicke model.  It can naturally arise in semiconductor quantum dot
lattices (see for e.g.~\cite{Raini_superfluor_quantumdots_Nature}),
where each quantum dot can have a varied orientation relative to
propagating electric fields, yet due to the small structure all dots
effectively radiate into a single mode, causing superradiance. One can
also engineer controlled realisations, by transforming a random
spatial distribution of atoms within an optical cavity
\cite{dimer2007proposed} relative to a varying electric field
amplitude into a distribution of couplings. Other possibilities
include ultra-cold molecules whose fixed transition dipoles are
randomly oriented with respect to the cavity field direction.
	
The organization of the article is as follows. In the next section we
will discuss the system Hamiltonian for the disordered Dicke model. In
section~\ref{sec_3} and \ref{sec_4} we present our results regarding
the two types of phase transitions: QPT and TPT.  In
section~\ref{sec_5} we outline several possible experimental
realisations. Finally in section~\ref{sec_6} we provide a summary of
our work.

\section{Model Hamiltonian and quantifiers}\label{sec_2} 

In Fig.~\ref{fig:schematic} we show a schematic of the disordered
Dicke model. The Hamiltonian consists of a single-mode bosonic field
coupled to $N$ atoms with a coupling strength that is modeled as a
random variable. The Hamiltonian can be written as
\begin{eqnarray}
  H = \omega a^{\dagger}a + \frac{\omega_0}{2}\sum_{i=1}^{N}\sigma_z^{(i)} + \frac{1}{\sqrt{N}}( a^{\dagger} + a )\sum_{i=1}^{N}g_i\sigma_x^{(i)},
  \label{eqn:H_DDM}
\end{eqnarray}
where the operators $a$ and $a^{\dagger}$ are the bosonic annihilation
and creation operators respectively, following the commutation
relation $[a, a^{\dagger}] = 1$ and
$J_{x,z}=\sum_{i=1}^{N}\frac{1}{2}\sigma_{ x,z}^{(i)}$ are the angular
momentum operators of a pseudospin with length $j$, composed of $N=2j$
spin-$\frac{1}{2}$ atoms described by Pauli matrices
$\sigma_{x,z}^{(i)}$ acting on site $i$. Here, the $g_i$'s are random
numbers drawn from two types of distributions. In the first
distribution, the $g_i$'s are drawn from a uniform unit box
distribution with finite width ($2\epsilon$) 
and height $A$ such that
$2\epsilon A = 1$. The parameters $\epsilon$ and $A$ are chosen so
that $2\epsilon = (\mu + \epsilon) - (\mu - \epsilon)$, $\epsilon =
\sqrt{3}\sigma$ and hence $A = \frac{1}{2\sqrt{3}\sigma}$ where $\mu$
and $\sigma$ are the mean and the standard deviation. 
In the second distribution, we consider $g_i\propto\cos\theta_i$,
where $\theta_i$ are angles randomly drawn from a Gaussian
distribution $p(\theta)\sim
\exp([-(\theta-\theta_0)^2/\sigma_\theta^2])$. Both can be engineered
e.g.~in optical cavities as sketched in \fref{fig:schematic} and
discussed in \sref{sec_5}. Due to the disorder in the coupling
strengths, $J^2$ is not a conserved quantity of the
Hamiltonian~\ref{eqn:H_DDM} and hence we have to consider all possible
spin configurations. For $N$ spins, the corresponding dimension of the
spin sub-space is $2^N$ and the bosonic sub-space dimension is
$n_{\text{max}} + 1$, where $n_{\text{max}}$ is the maximal occupation
we allow for the bosonic field.  Hence the total Hilbert space
dimension for our numerical calculations is $N_D=2^N( n_{\text{max}} +
1 )$.
    
In the next sections we explore the QPT and TPT separately, based on
the properties of eigenvalues and eigenstates of \eref{eqn:H_DDM}.  We
study useful quantifiers such as ground state energy, average boson
number and mutual information between two spins. For a mixed state
(like a temperature equilibriated state), the mutual information has
been shown~\cite{das2022revisiting} to be an appropriate quantity,
although it contains both quantum and classical correlations.  We
shall use the mutual information~\cite{vedral2003classical,
  divincenzo2004locking, adesso2010quantum, maziero2010quantum,
  das2022revisiting} between two spins from a mixed density matrix.
Defining the reduced density matrices of any two selected spins to be
$\rho_1$ and $\rho_2$ and the reduced density matrix corresponding to
the two-spin state to be $\rho_{12}$, the mutual information between
the two spins can be computed using the relation:
\begin{equation}
  I_{12} = S_1 + S_2 - S_{12},
\end{equation}
where $S_{1,2}= -Tr(\rho_{1,2} \ln(\rho_{1,2}))$, $S_{12} =
-Tr(\rho_{12}\ln(\rho_{12}))$ are the corresponding von Neumann
entropies. Since we will be interested in $I_{12}$ at finite
temperature, we will first construct the total thermal density matrix:
$\rho_{\text{Th}}=e^{-\frac{\mathcal{H}}{k_B T}}$, and then trace over
the bosonic subspace and the remaining $(N-2)$ or $(N-1)$ spins. Since
we will average over the disorder, it does not matter which two spins
are considered for the purpose of computing mutual
information. Another useful observable that we use to study the QPT is
the average boson number $\langle a^{\dagger} a\rangle$ evaluated in
the interacting ground state.

\section{Quantum phase transition}\label{sec_3}
It is well known that in the thermodynamic limit (when the atom number
$N\to \infty$), the usual Dicke model exhibits a quantum phase
transition~\cite{lambert2004entanglement} from the normal phase to the
super-radiant phase at some critical coupling strength $g_c$. In the
disordered Dicke model, if we fix the mean of the coupling strength at
a sufficiently low value and vary the standard deviation ($\sigma$) we
see a similar QPT. The QPT here is studied with the aid of the
disorder-averged energy and average boson number in the ground
state. In Figs.~\ref{fig:qpt_dm} and \ref{fig:qpt2_dm} we show these
properties in the ground-state of the disordered Dicke model,
considering two types of distributions as discussed in the previous
section.

What we will empirically show now, is that much of the behavior of the disordered 
Dicke model can be understood by averaging known results for the 
disorder-free (clean) model. This is not clear a-priori, since all 
the two-level systems in the disordered model couple to the same 
bosonic mode, and thus get coupled. For the clean Dicke model, 
Emary et al~\cite{emary2003quantum} have derived analytical results 
for the ground state energy:
\begin{align}
  E_G &= 
  \begin{cases}
    -\frac{N \omega_0}{2},              & g < g_c\\
    - \frac{N \omega_0}{4}\left[ \frac{g^2}{g_c^2} + \frac{g_c^2}{g^2}\right] ,              & g > g_c
  \end{cases}
  \label{eqn:E_gs_dm}
\end{align}
and the average boson number in the cavity:
\begin{align}
  \langle a^{\dagger}a \rangle &= 
  \begin{cases}
    0,              & g < g_c\\
    \frac{N}{\omega^2}\left[ g^2 - \frac{g_c^4}{g^2} \right]              & g > g_c
  \end{cases}
  \label{eqn:boson_dm}
\end{align}
where $g_c$ is the critical value of the coupling in the absence of
disorder. We will make use of the above results and integrate over
the coupling strength distribution to obtain approximate analytical results for the
disordered Dicke model. We denote the disorder-averaged value of an
observable $O$ as $\overline{O}$:
\begin{equation}
  \overline{O} = \int_{x_1}^{x_2} P(g)O(g) dg,
\end{equation}
where $P(g)$ is the distribution of the disorder and the limits of
integration $x_1$ and $x_2$ have to be chosen appropriately according
to the observable and the distribution being considered.

\subsection{Uniform distribution}\label{sec_3a}
\begin{figure}[t]
  \subfigure{\includegraphics[width=0.235\textwidth]{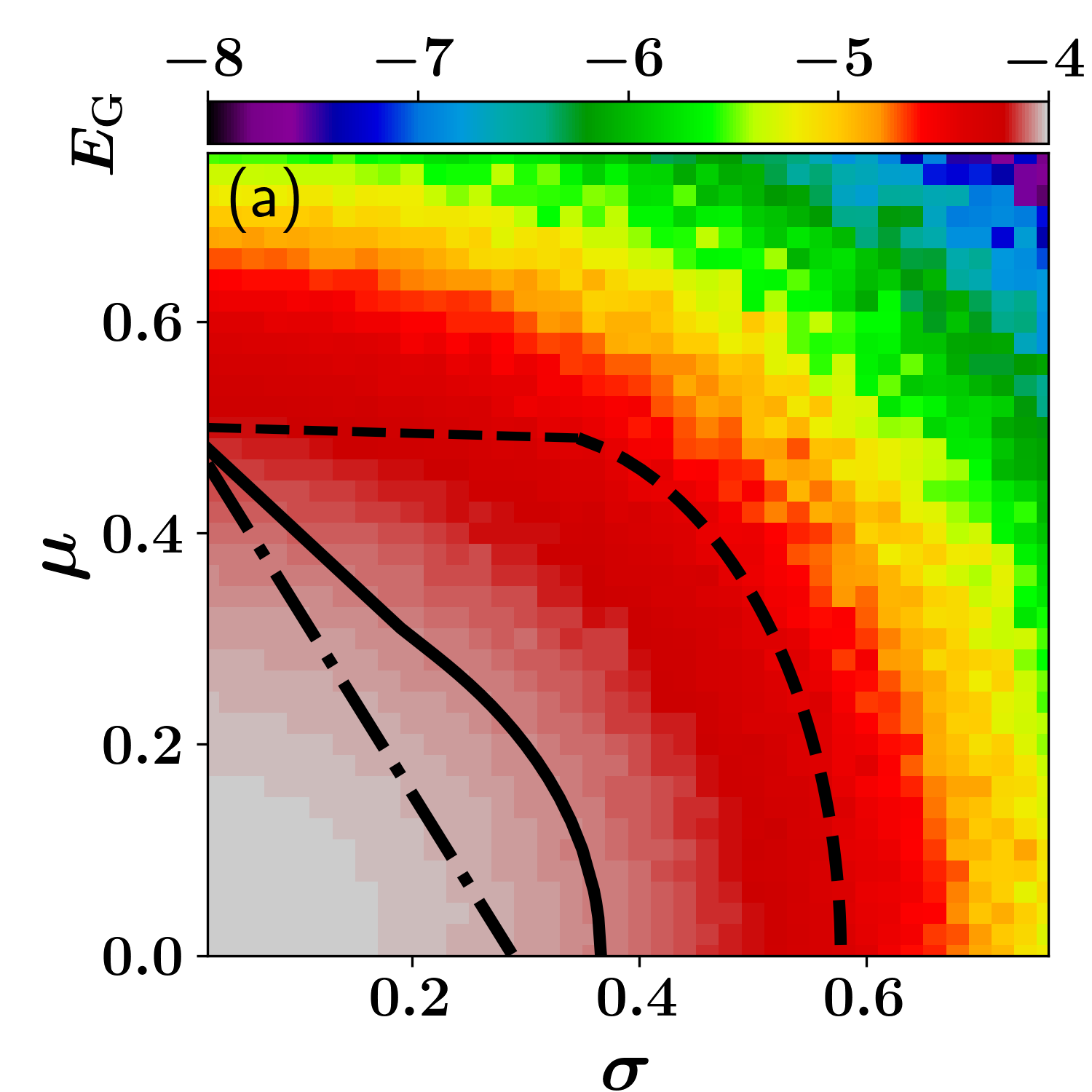}\label{fig:boson_qpt}}
  \subfigure{\includegraphics[width=0.235\textwidth]{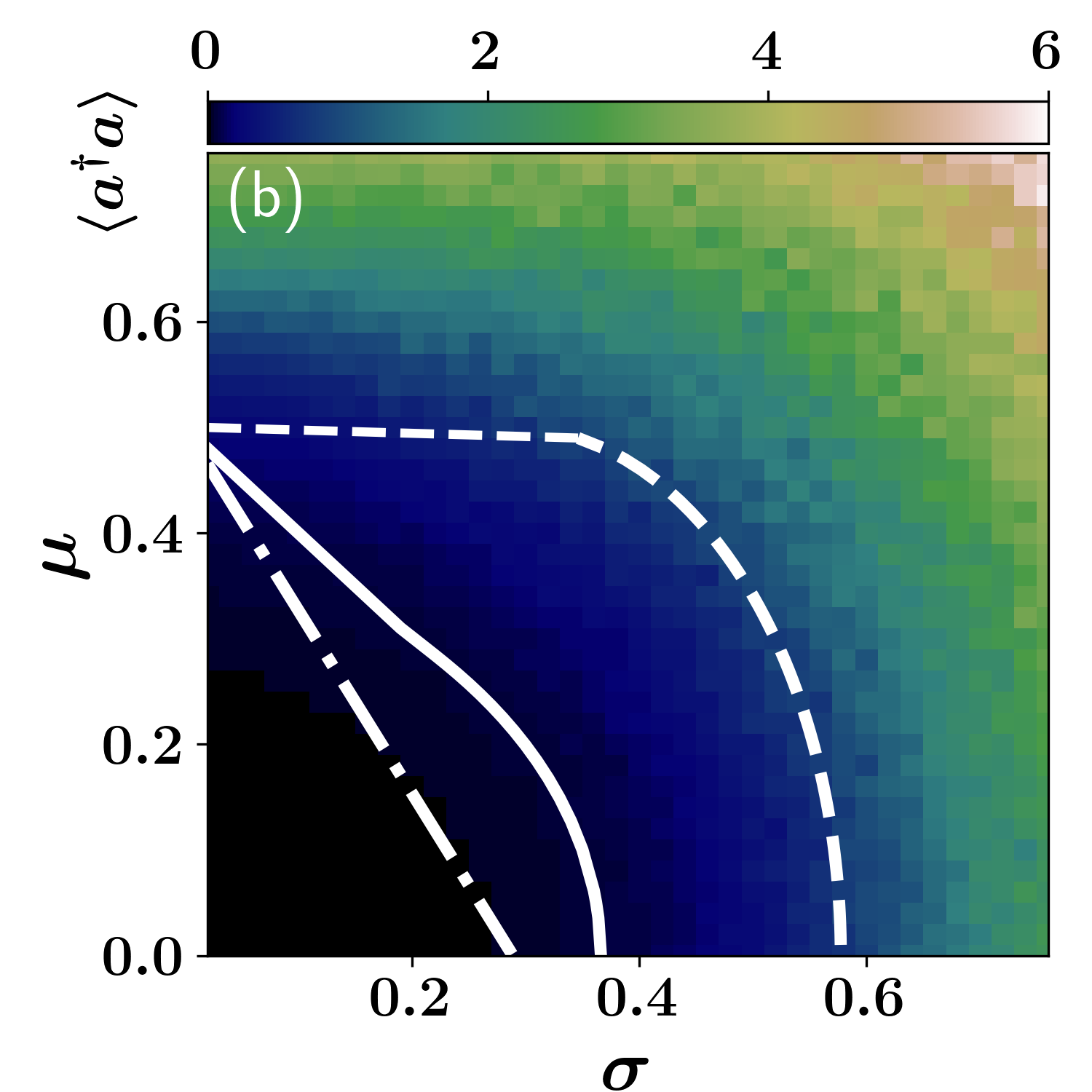}\label{fig:boson_qpt}}
  \caption{Phase diagram of the disordered Dicke model with uniform
    coupling distribution (\eref{eqn:uniform}). To map it out, we show
    (a) the ground state energy $E_{\text{G}}$ and (b) the average
    boson number, $\langle a^{\dagger}a\rangle$ wrt.~the ground state,
    as a function of the standard deviation $\sigma$ and the mean
    $\mu$ of the coupling parameters $g_i$. We consider the resonant
    case: $\omega = \omega_0 = 1$, take the average over $120$
    realizations, and fix the atom number to be $N=8$, the bosonic
    cut-off to be $n_{\text{max}}=40$.}
  \label{fig:qpt_dm}
\end{figure}

In the first scenario the coupling $g$ is drawn from a uniform
distribution:
\begin{equation}
  P_{u}(g) = \begin{cases}\frac{1}{2\sqrt{3}\sigma} \quad \textrm{if}\quad \mu-\sqrt{3}\sigma < g < \mu+\sqrt{3}\sigma\\
    0 \quad\quad\quad \textrm{otherwise}
    \end{cases}
\label{eqn:uniform}     
\end{equation}
with mean $\mu$ and standard deviation $\sigma$.

The disorder-averaged ground state energy and average boson number, 
are found from the integrals:
\begin{align}
  \overline{E_G} &= \int_{x_1}^{x_2} P_{u}(g) E_G dg,\label{eqn:average1}\\
  \overline{\langle a^{\dagger}a\rangle} &= \int_{x_1}^{x_2} P_{u}(g) \langle a^{\dagger}a\rangle dg,  
  \label{eqn:average2}
\end{align}
where $E_G$ is given in Eqn.~\ref{eqn:E_gs_dm}, $\langle
a^{\dagger}a\rangle$ is given in Eq.~\ref{eqn:boson_dm}, and we use the overline to denote the disorder average.
 The lower and upper limits of the box distribution are: $x_1
= \mu - \sqrt{3}\sigma$ and $x_2 = \mu + \sqrt{3}\sigma$
respectively and we consider $\mu$ and $\sigma$ to be in the
range: $[0,1]$.

For the NP ($|g|\leq g_c$) we find (Appendix \ref{app_uniform_calculations}):
\begin{align}
  \overline{E_G} &=  -\frac{N\omega_0}{2},\\
  \overline{\langle a^{\dagger}a\rangle} &= 0.
\end{align} 
On the other hand, for the SP ($|g| > g_c$), we have:
\begin{align}
  \overline{E_G} &= - \frac{N}{2\sqrt{3}\sigma} \left[ \frac{x_2^3}{3} - \frac{g_c^4}{x_2} + \frac{2g_c}{3} \right],\\
  \overline{\langle a^{\dagger}a\rangle} &= \frac{N}{2\sqrt{3}\sigma\omega^2} \left[ \frac{x_2^3}{3} + \frac{g_c^4}{x_2} - \frac{4g_c^3}{3} \right].
\end{align}
Taylor expanding around the critical point $g_c$
and considering only the dominant terms, we have:
\begin{eqnarray}
  \overline{E_G} &\approx& - \frac{N\omega_0}{2} - \frac{AN \omega_0}{2} ( x_2 - g_c ) - 1.33 AN \omega_0 ( x_2 - g_c )^3,\nonumber\\
  \label{eqn:Egs_avg_dm}
\end{eqnarray}
\begin{eqnarray}
  \overline{\langle a^{\dagger}a\rangle} &\approx& \frac{A N}{\omega^2} ( x_2 - g_c )^2 - \frac{0.667 A N}{\omega^2} ( x_2 - g_c )^3,
  \label{eqn:boson_avg_dm}
\end{eqnarray}
where $A = \frac{1}{2\sqrt{3}\sigma}$ and $x_2$ is the upper limit of
the integration 
$\mu+\sqrt{3}\sigma$. At the
critical point the ground state energy is $-\frac{N\omega_0}{2}$ and
the average boson number is zero, hence we have a relation for the
critical line as a function of $\mu$ and $\sigma$:
\begin{eqnarray}
  \mu + \sqrt{3}\sigma = \frac{1}{2}.
  \label{eqn:critical_dm}
\end{eqnarray}

Fig.~\ref{fig:qpt_dm}(a) shows the numerical value of the ground state
energy $E_{\text{G}}$ of the system as a function of the standard
deviation ($\sigma$) and mean ($\mu$) of the coupling parameter. Our
goal is to check the validity of the equation for the critical line
marking the QPT (Eq.~\ref{eqn:critical_dm}) numerically. In this
figure the white/pink color indicates the normal phase where the
ground state energy is large and constant, $ E_{\text{G}} =
-\frac{N\omega_0}{2}$ and the other colors represent the super-radiant
phase where $E_{\text{G}}$ is decreasing.  Similarly
Fig.~\ref{fig:qpt_dm}(b) shows the average boson number in the ground
state of the disordered Dicke model. In the normal phase
$\overline{\langle a^{\dagger}a\rangle} \approx 0$ (black color),
i.e.~there are no excitations in the bosonic mode whereas in the
super-radiant phase $\overline{\langle a^{\dagger}a\rangle}$ is finite
(other colors), which indicates a macroscopic excitation of the
bosonic mode. The dash-dotted line here represents the quantum
critical line which is given in Eq.~\ref{eqn:critical_dm} and our
numerical data already roughly agrees with this linear relation. It is
remarkable that the formula describes the numerical data this well,
despite the coarse approximation of just disorder-averaging the clean
Dicke model results.  Around the critical line the expectation value
(with respect to the uniform disorder) of the ground state energy and
the average boson number can be represented by the simpler Taylor
series in Eq.~\ref{eqn:Egs_avg_dm} and Eq.~\ref{eqn:boson_avg_dm}
respectively.  It is clear that for $\mu=0$, the standard deviation of
the disordered Dicke model plays the same role as the coupling
parameter $g$ in the usual Dicke model and the critical point is
$\sigma_c = \frac{g_c}{\sqrt{3}}$ within this crude approximation.

\begin{figure}[t]
  \includegraphics[width=0.45\textwidth]{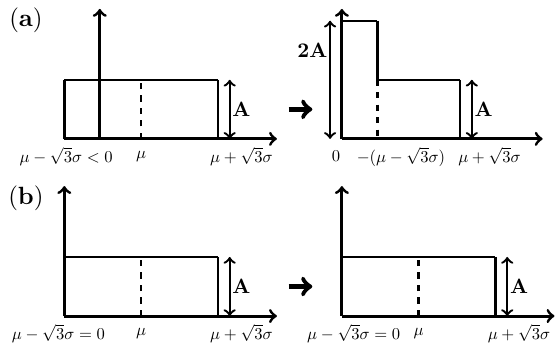}\label{fig:boson}
  \caption{Effective distribution containing only positive coupling
    strengths.  The left panel denotes the original distribution,
    whereas the right panel shows the effective distribution where
    only the absolute values of the coupling strengths are
    considered. (a) For $\mu-\sqrt{3}\sigma<0$, the effective mean of
    the coupling strength is $\langle g\rangle =
    \frac{\mu+3\sigma^2}{2\sqrt{3}\sigma}$ and the effective variance
    $\langle g^2\rangle = \mu^2 + \sigma^2$ hence the effective
    standard deviation is $\sqrt{\mu^2 + \sigma^2 -
      \frac{(\mu+3\sigma^2)^2}{12\sigma^2}}$.  (b) For
    $\mu-\sqrt{3}\sigma\geq 0$ the distribution remain unchanged: the
    mean and the standard deviation of $g$ are $\mu$ and $\sigma$.
    For both the cases $A=\frac{1}{2\sqrt{3}\sigma}$.  }
  \label{fig:uni_dis}
\end{figure}
The line that separates the NP and the SP in Fig.~\ref{fig:qpt_dm} can also
be obtained approximately with the aid of a heuristic argument that
exploits a symmetry of the Hamiltonian. We observe that the
Hamiltonian in Eqn.~\ref{eqn:H_DDM} has the same eigenvalues as one in which any one of
the couplings $g_i$ is changed to $-g_{i}$. In other words, the eigenvalues of $H(\{g_{j},j\neq i\},g_i)$ and 
$H(\{g_{j},j\neq i\},-g_i)$ are the same. This is a direct consequence of the fact that
\begin{equation}
  H(\{g_j,j\neq i\},-g_i) = \sigma_i^z H(\{g_j,j\neq i\},g_i)\sigma_i^z.
\end{equation}
Thus when the transformation $T = \sigma_i^{z}$ is applied on any
eigenstate of the Hamiltonian $H(\{g_{j},j\neq i\},g_i)$, we would get
an eigenstate of the Hamiltonian $H(\{g_{j},j\neq i\},-g_i)$ with the
same eigenvalue. This argument 
naturally extends to the case when
multiple $g_i$'s undergo a sign change. Hence we can consider a
scenario where all the coupling strengths are made positive, i.e.~if
there are any negative coupling strengths, we simply take their
absolute values. Hence when the lower limit of the uniform
distribution $\mu - \sqrt{3}\sigma < 0$ the effective distribution is:
\begin{equation}
  P_{\textrm{eff}}(g) = \begin{cases}\frac{1}{\sqrt{3}\sigma} \quad\quad \textrm{if}\quad 0 < g < -(\mu-\sqrt{3}\sigma)\\
    \frac{1}{2\sqrt{3}\sigma} \quad \textrm{if} -(\mu-\sqrt{3}\sigma) < g < (\mu+\sqrt{3}\sigma).
    \end{cases}
\label{eqn:uniform2}     
\end{equation}
as shown in Fig.~\ref{fig:uni_dis}(a). The effective distribution in
this case yields a mean value of $\langle g\rangle =
\frac{\mu+3\sigma^2}{2\sqrt{3}\sigma}$ and a variance of $\langle
g^2\rangle = \mu^2 + \sigma^2$ which in turn corresponds to a standard
deviation of: $\sqrt{\mu^2 + \sigma^2 -
  \frac{(\mu+3\sigma^2)^2}{12\sigma^2}}$.  If the lower limit of the
distribution $\mu-\sqrt{3}\sigma\geq 0$, the effective distribution
remains identical to the original one and its mean and standard
deviation remain unchanged as $\mu$ and $\sigma$
(Fig.~\ref{fig:uni_dis}(b)).

To identify the phase transition line heuristically, we argue as
follows. We would expect that as more and more of the couplings $g_i$ are
drawn 
above $g_c$, we would see increasingly dominant effects characteristic of the SP. 
A coarse way to identify this would be to simply demand that the right most
edge of the effective distribution (Eqn.~\ref{eqn:uniform2}) must be above the critical coupling $g_c=\frac{1}{2}$,
i.e.
\begin{equation}
  \mu + \sqrt{3}\sigma = \frac{1}{2},
\end{equation}  
which is nothing but the crude approximation 
Eqn.~\ref{eqn:critical_dm} 
and dot-dashed line in
Fig.~\ref{fig:qpt_dm}. For a refined result, we 
demand that the mean of
the effective distribution (Eqn.~\ref{eqn:uniform2}) must reach above $g_c$ 
\begin{eqnarray}
  \frac{\mu+3\sigma^2}{2\sqrt{3}\sigma} = \frac{1}{2},\hspace{4mm} \mu<\sqrt{3}\sigma\nonumber\\
  \mu  = \frac{1}{2}, \hspace{3mm} \mu\geq\sqrt{3}\sigma.
\end{eqnarray}
This is shown by the dashed line in Fig.~\ref{fig:qpt_dm}. A less stringent condition is to
demand that the mean plus one standard deviation of the effective distribution (Eqn.~\ref{eqn:uniform2}) must reach above $g_c$ 
\begin{eqnarray}
\frac{\mu+3\sigma^2}{2\sqrt{3}\sigma} + \sqrt{\mu^2 + \sigma^2 - \frac{(\mu+3\sigma^2)^2}{12\sigma^2}} = 0.5,\hspace{4mm} \mu<\sqrt{3}\sigma\nonumber\\
\mu + \sigma = 0.5, \hspace{3mm} \mu\geq\sqrt{3}\sigma.\nonumber\\
\end{eqnarray}
This is shown by the solid white line in Fig.~\ref{fig:qpt_dm} and appears
to be closest to the actual line of separation between the SP and NP.

\begin{figure}[t]
  \subfigure{\includegraphics[width=0.235\textwidth]{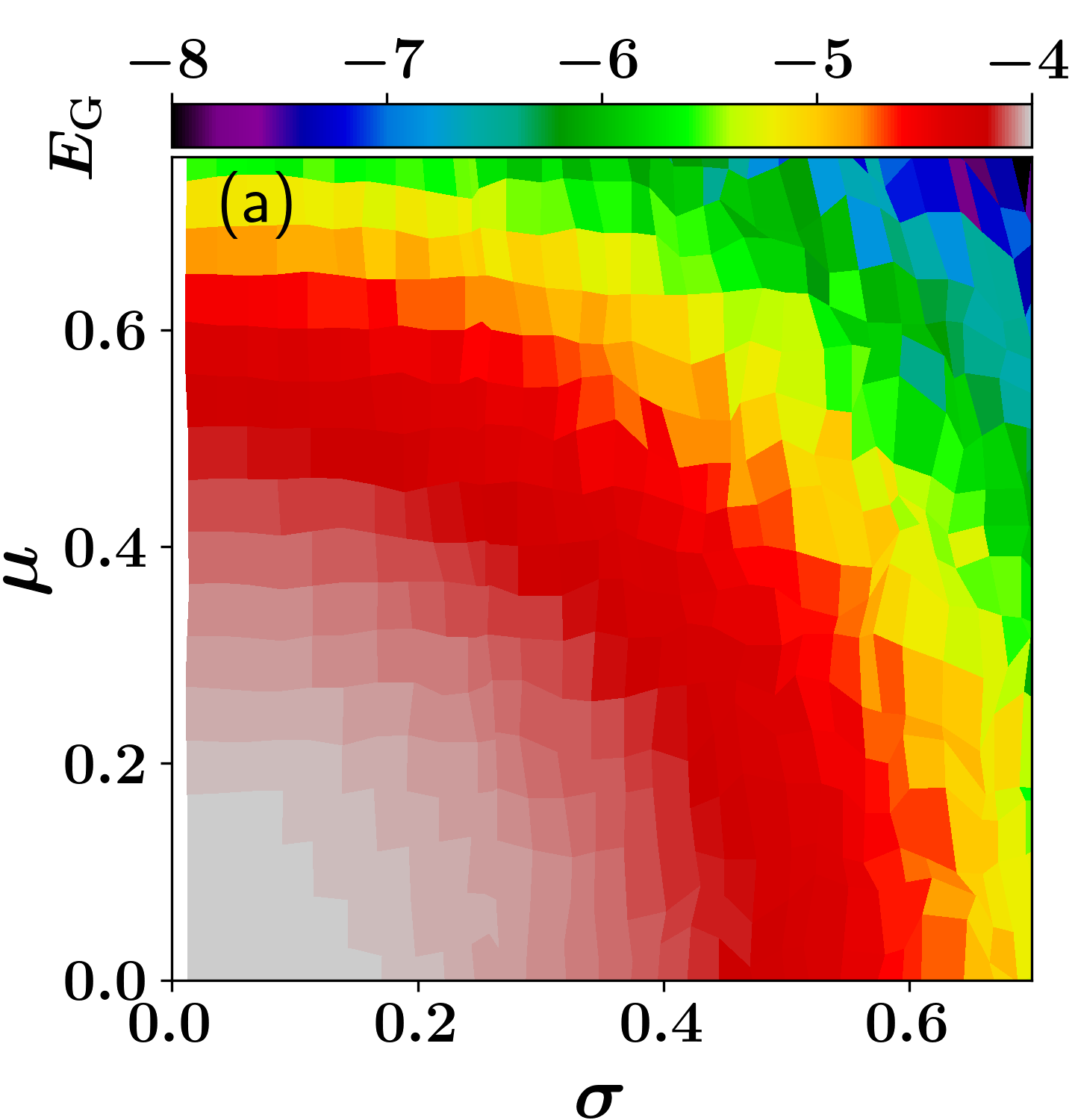}\label{fig:boson_qpt}}
  \subfigure{\includegraphics[width=0.235\textwidth]{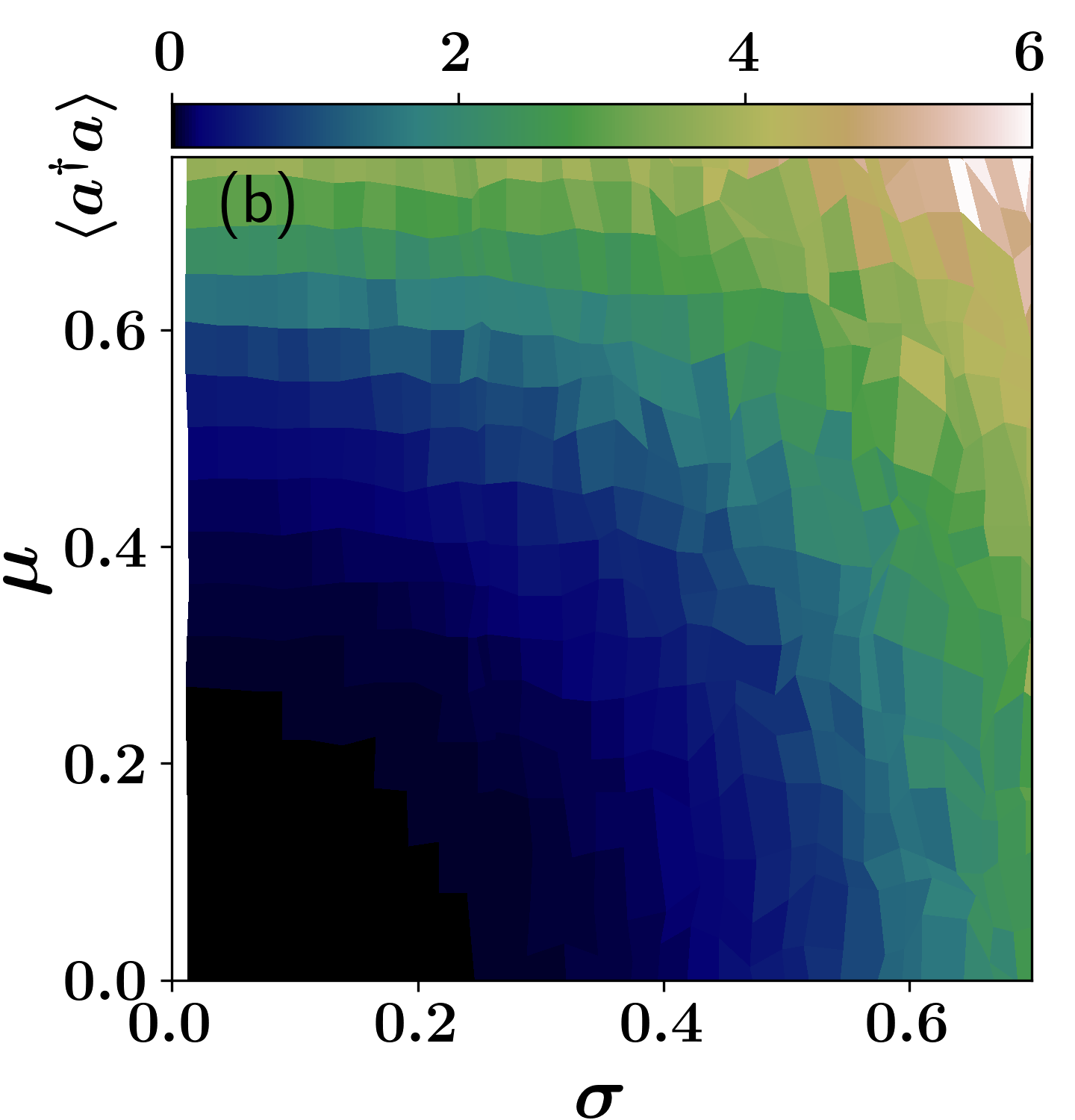}\label{fig:boson_qpt}}
  \caption{Phase diagram of the disordered Dicke model where $g_i=2\cos\theta_i$, $\theta_i$ are angles
    randomly drawn from a Gaussian distribution with mean $\theta_0$
    and standard deviation $\sigma_\theta$.  To map it out, we show
    (a) the ground state energy $E_{\text{G}}$ and (b) the average
    boson number, $\langle a^{\dagger}a\rangle$ wrt. the ground state,
    as a function of the standard deviation $\sigma$ and the mean
    $\mu$ of the coupling parameters $g_i$. We consider the resonant
    case: $\omega = \omega_0 = 1$ and take the average over $200$
    realizations, and fix the atom number to be $N=8$, and the bosonic
    cut-off to be $n_{\text{max}}=40$.}
  \label{fig:qpt2_dm}
\end{figure}

\subsection{Gaussian distribution}\label{sec_3c}
To demonstrate the robustness of our results to variations of the
detailed shape of the probability distribution for the coupling, we
now consider a second case. The angle $\theta$ is drawn from the
Gaussian distribution:
\begin{equation}
P(\theta) \propto e^{-(\theta-\theta_0)^2/\sigma_\theta^2},
\label{angle_dist}
\end{equation}
where $\theta_0\in [0,\pi]$ is the mean and $\sigma_\theta\in [0,\frac{\pi}{4}]$ 
is the standard deviation of $\theta$ and the disordered coupling strength 
for the $i^{\text{th}}$ spin is then taken as:
\begin{equation}
g_i = 2\cos\theta_i.
\label{angle_to_coupling}
\end{equation} 
Here, we numerically calculate the mean and the standard deviation of
$g$: $\mu = \langle g\rangle = \frac{1}{N}\sum_{i=1}^N g_i$ and
$\sigma = \sqrt{\langle g^2\rangle - \langle g\rangle^2}$, to obtain
characteristics of the distribution that are easily comparable with
the previous section.  The ground state energy $E_{\text{G}}$ and the
average boson number for this distribution are shown in
Fig.~\ref{fig:qpt2_dm} and we see a behavior similar to the uniform
distribution used in Fig.~\ref{fig:qpt_dm}.

\section{Thermal phase transition}\label{sec_4}
\begin{figure*}[t]
  \subfigure{\includegraphics[width=0.235\textwidth]{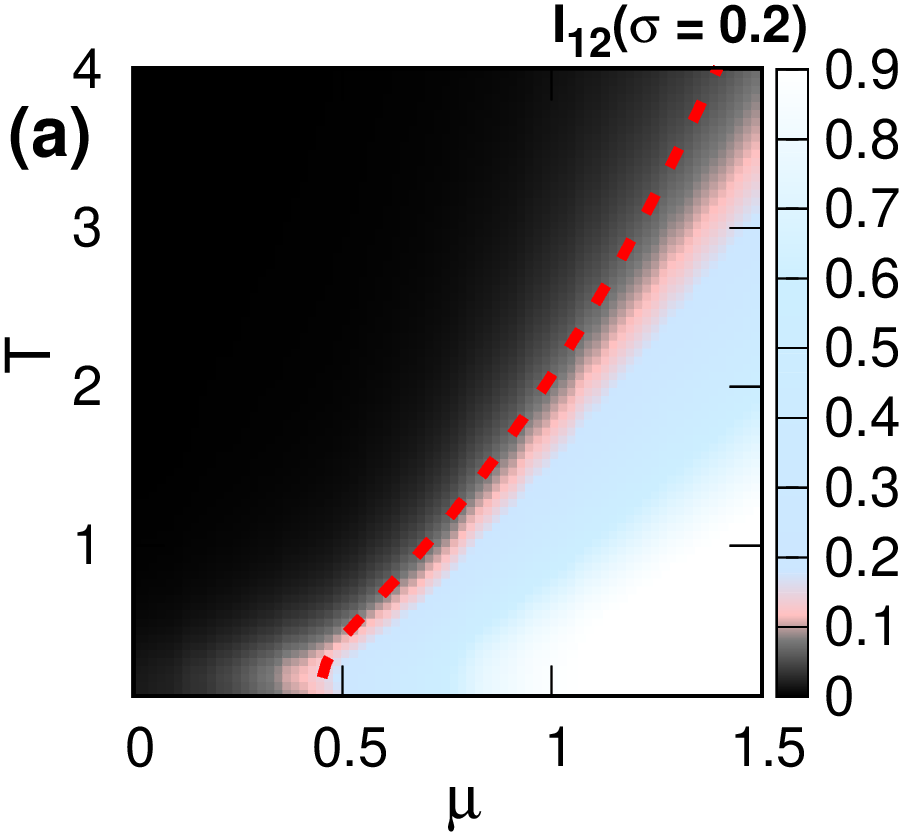}\label{fig:MI_sig0pt2}}
  \subfigure{\includegraphics[width=0.235\textwidth]{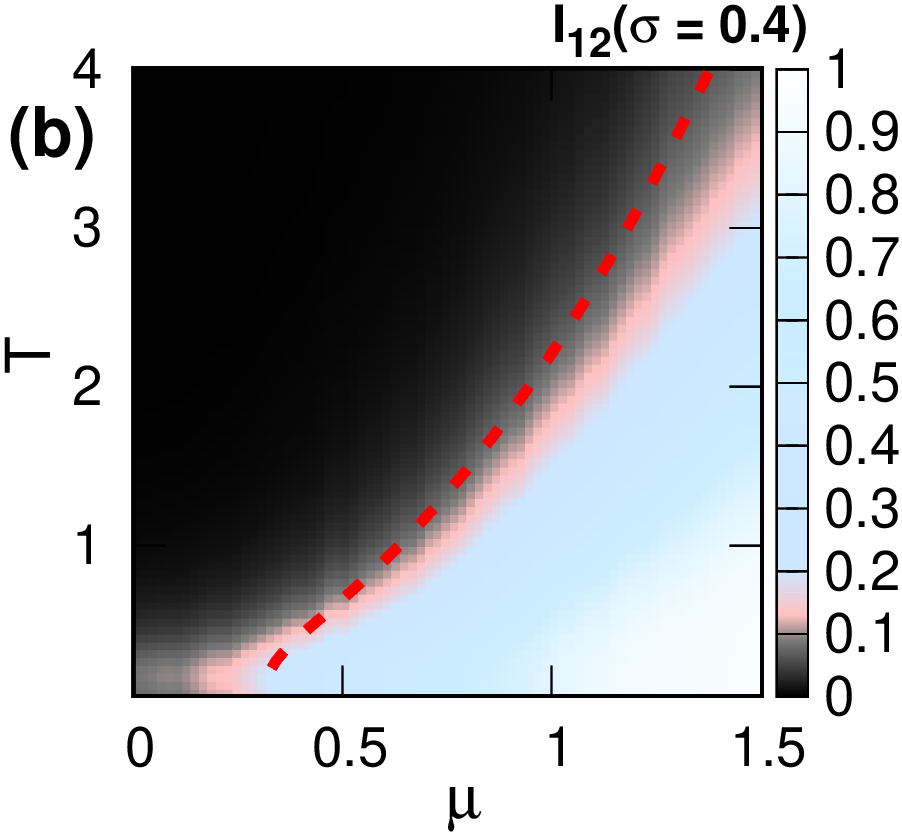}\label{fig:MI_sig0pt4}}
  \subfigure{\includegraphics[width=0.235\textwidth]{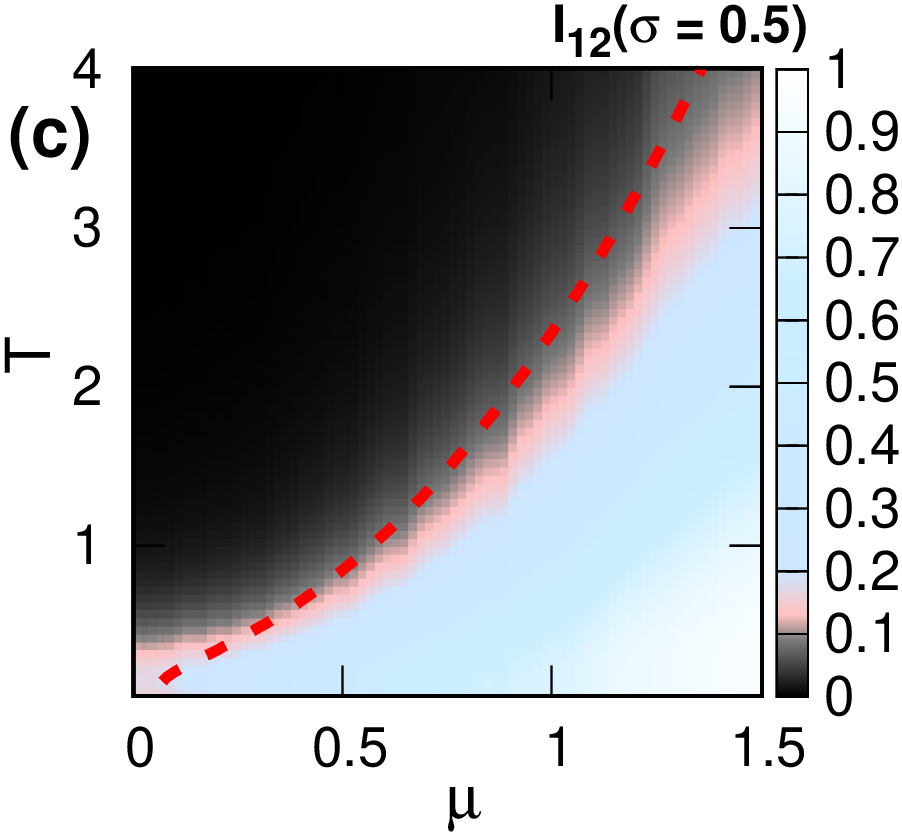}\label{fig:MI_sig0pt5}}
  \subfigure{\includegraphics[width=0.235\textwidth]{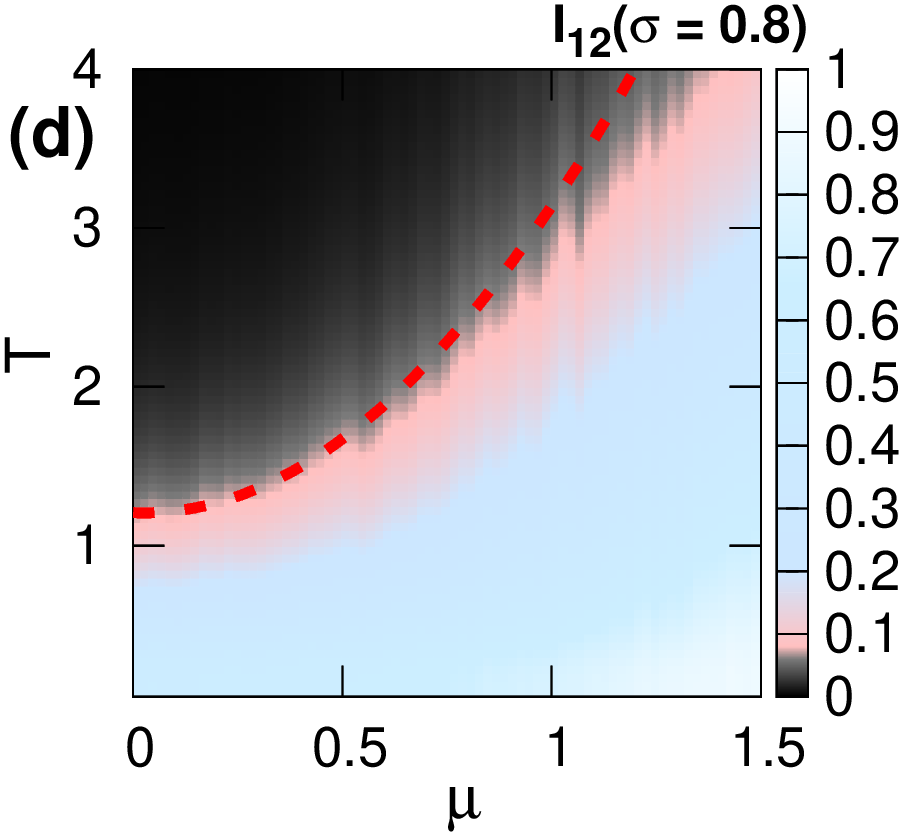}\label{fig:MI_sig1pt0}}     
  \caption{ Thermal phase diagrams of the disordered Dicke model,
    based on the mutual information between two spins. Axes are the
    temperature $T$ and mean coupling strength $\mu=\langle g\rangle$,
    for (a) $\sigma=0.2$, (b) $\sigma = 0.4$, (c) $\sigma = 0.5$, (d)
    $\sigma=0.8$. The couplings $g$ are drawn from a random uniform
    distribution with finite mean $\mu$ and standard deviation
    $\sigma$ (see Eq.~\ref{eqn:uniform}). The number of atoms is $N=6$
    and we choose the bosonic cut-off as $n_{\text{max}}=20$. We take
    the average over $824$ realizations of $g$ for each $\sigma$.}
  \label{fig:MI}
\end{figure*}    
Moving from the quantum to the thermal phase transition, in this
section we derive an analytical expression for the critical
temperature for the disordered Dicke model building on previous
results~\cite{wang1973phase, hioe1973phase} for the clean Dicke
model. We start by rewriting the system Hamiltonian for the disordered
Dicke model as:
\begin{align}
  \tilde{\mathcal{H}} &= \frac{\mathcal{H}}{\omega}\nonumber\\
  &= a^{\dagger}a + \sum_{j=1}^{N}\frac{\epsilon}{2}\sigma_j^z + \frac{1}{\sqrt{N}}( a + a^{\dagger} )\sum_{j=1}^{N}\lambda_j\sigma^x_j\\
  &= a^{\dagger}a + \sum_{j=1}^{N}h_j.
\end{align}
where $\epsilon = \frac{\omega_0}{\omega}$,
$\lambda_j=\frac{g_j}{\omega}$ and
\begin{align}
  h_j &= \frac{\epsilon}{2}\sigma_j^z + \frac{1}{\sqrt{N}}( a + a^{\dagger} )\lambda_j\sigma^x_j.
\end{align} 
Following Wang and Hieo~\cite{wang1973phase}, who studied the Dicke 
model within the rotating wave approximation, the partition 
function can be computed as:
\begin{align}
  Z(N,T) = \sum_{s_1,..., s_N=\pm 1}\int \frac{d^2\alpha}{\pi}\langle s_1...s_N\vert\langle\alpha\vert e^{-\beta\tilde{\mathcal{H}}}\vert\alpha\rangle\vert s_1...s_N\rangle\nonumber\\
  = \int\frac{d^2\alpha}{\pi}e^{-\beta\vert\alpha\vert^2}\Pi_{j=1,2,...,N}\langle s_j\vert e^{-\beta h_j}\vert s_j\rangle\nonumber\\
  = \int\frac{d^2\alpha}{\pi}e^{-\beta\vert\alpha\vert^2}\Pi_{j=1,2,...,N}\Big(2\cosh\Big[ \frac{\beta\epsilon}{2}\Big[ 1 + \frac{16{\lambda_j}^2 {\alpha}^2}{\epsilon^2 N} \Big]^{1/2} \Big]\Big).
  \label{eqn:partition_func}  
\end{align}
\begin{figure*}[t]
  \subfigure{\includegraphics[width=0.235\textwidth]{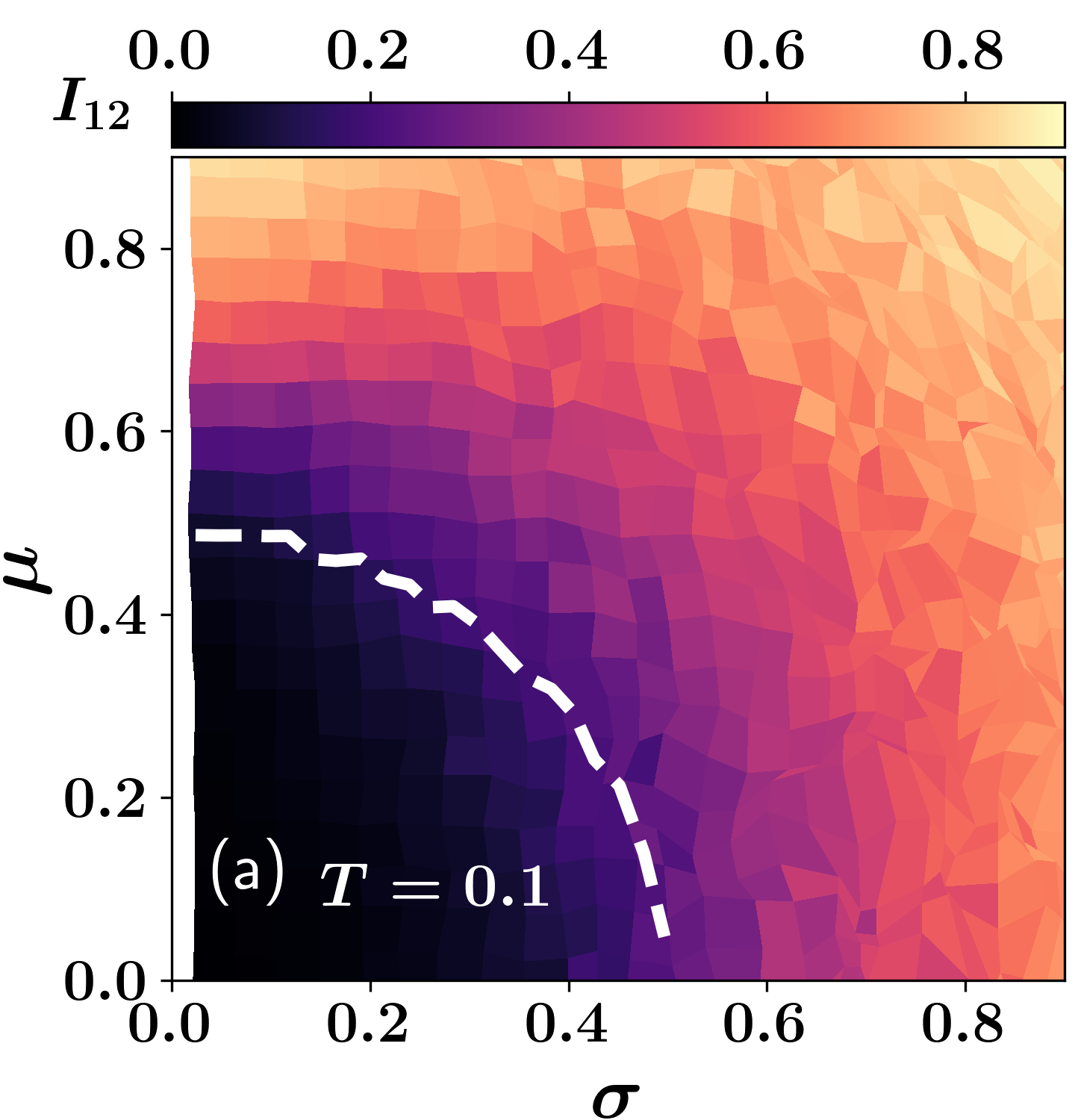}\label{fig:MI_sig0pt2}}
  \subfigure{\includegraphics[width=0.235\textwidth]{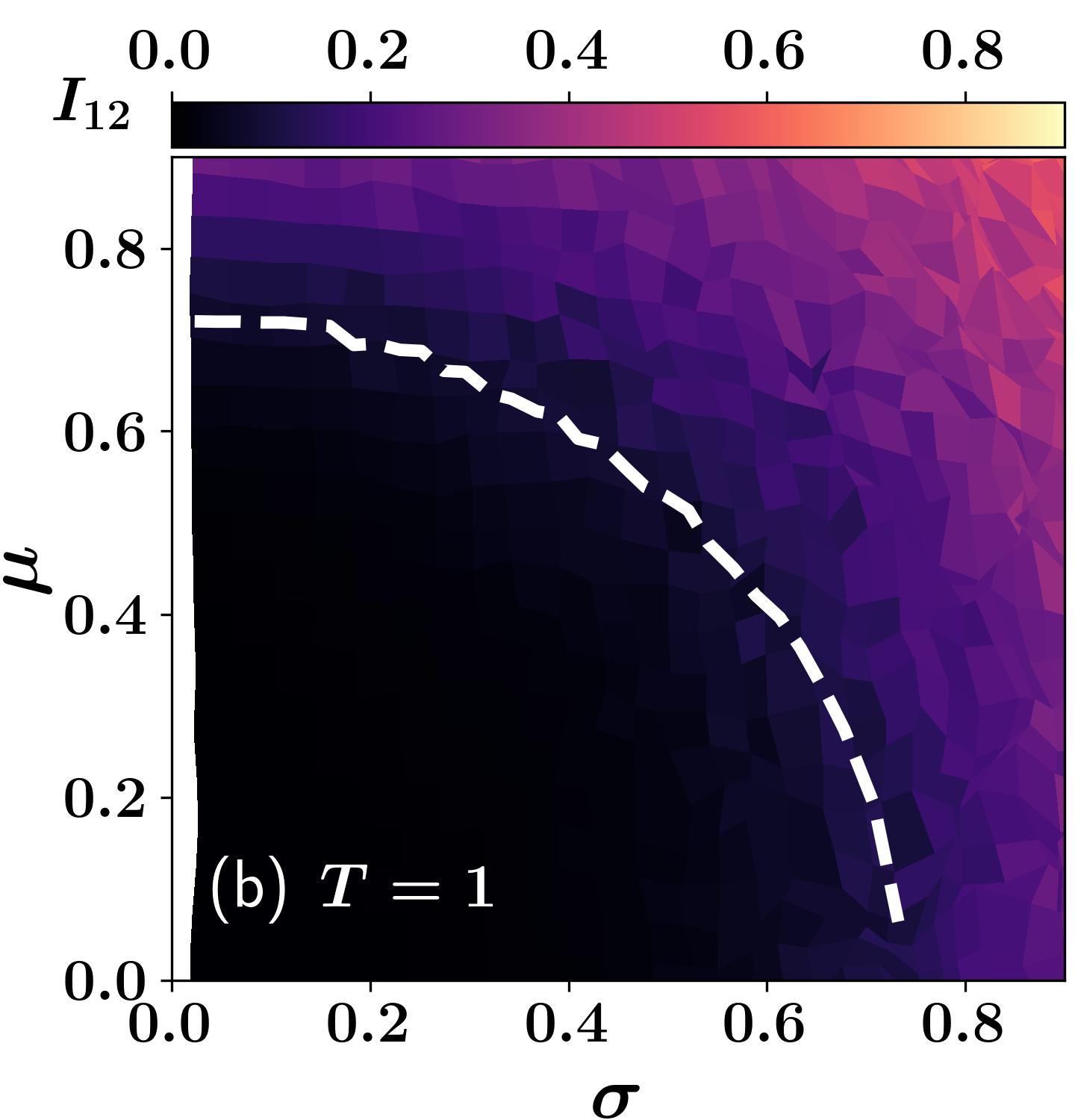}\label{fig:MI_sig0pt4}}
  \subfigure{\includegraphics[width=0.235\textwidth]{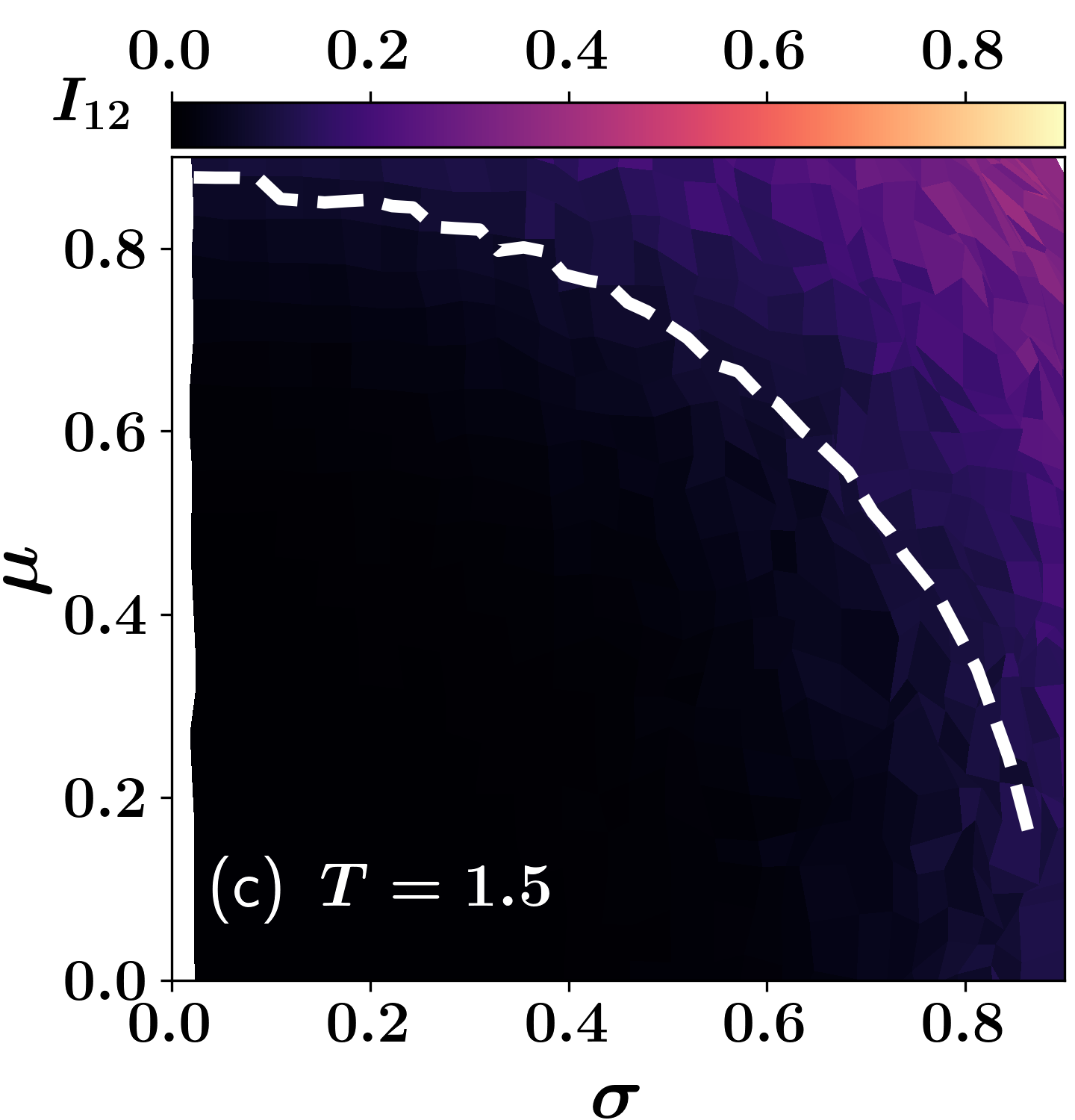}\label{fig:MI_sig0pt5}}
  \subfigure{\includegraphics[width=0.235\textwidth]{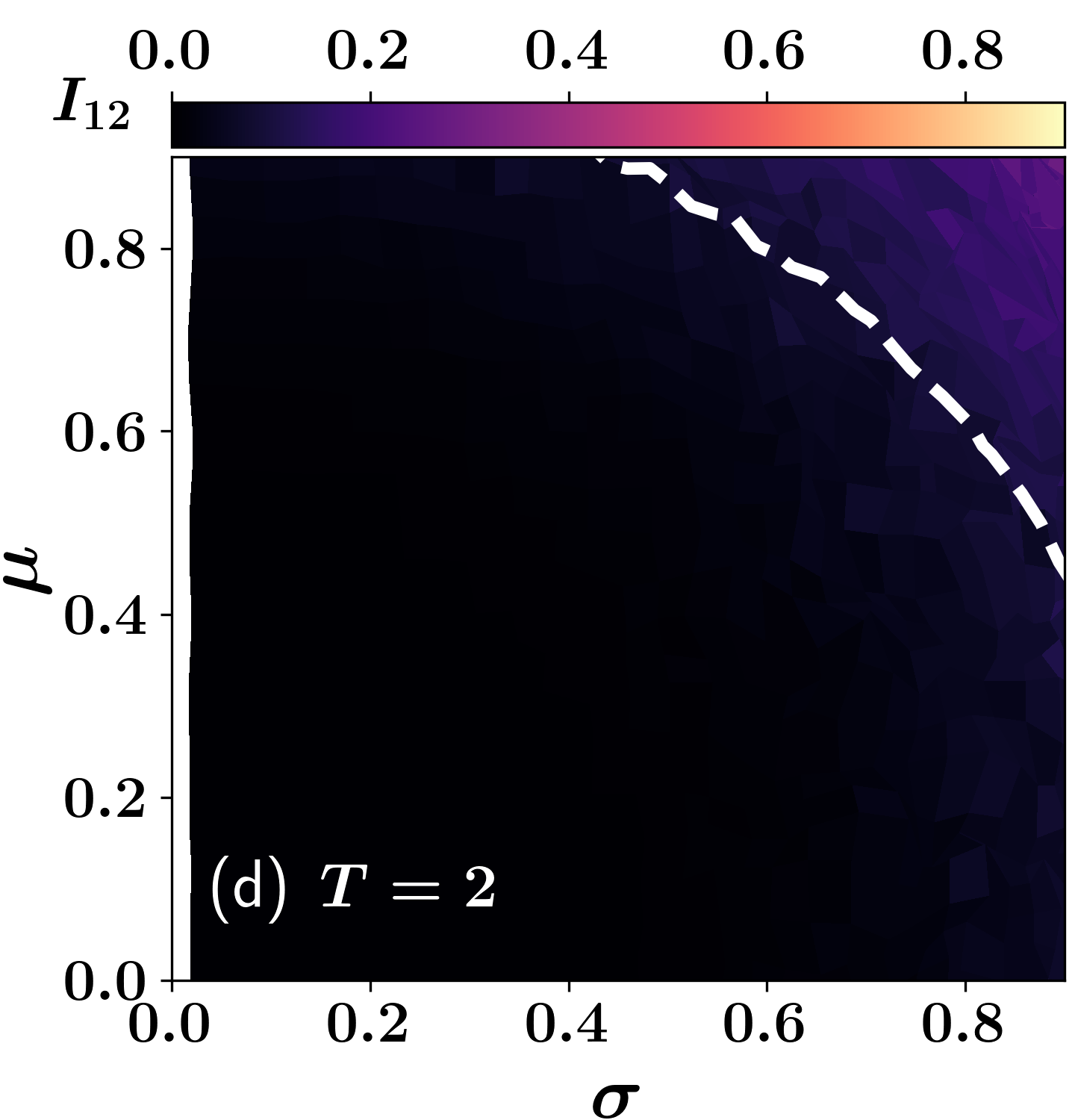}\label{fig:MI_sig1pt0}}
  
  \caption{ Thermal phase diagrams of the disordered Dicke model,
    based on the mutual information between two spins. Axes are the
    standard deviation $\sigma$ and mean coupling strength
    $\mu=\langle g\rangle$, for (a) $T = 0.1$, (b) $T = 1$, (c)
    $T=1.5$, (d) $T=2$.  The couplings $g_i = 2\cos\theta_i$, where
    $\theta_i$ are angles randomly drawn from a Gaussian distribution
    with mean $\theta_0$ and standard deviation $\sigma_\theta$. The
    number of atoms is $N=6$ and we choose the bosonic cut-off as
    $n_{\text{max}}=20$. We take the average over $96$ realizations of
    $g$ for each temperature.}
      \label{fig:MI2}
    \end{figure*}
    Here $|\alpha\rangle$ is a coherent state which satisfies the
    relation: $a|\alpha\rangle = \alpha|\alpha\rangle$ and $|
    s_1...s_N\rangle$ is the product basis for the spin subspace.  In
    polar coordinates the partition function becomes:
    \begin{widetext}
    \begin{align}
      Z(N, T) = \int_0^{\infty}r dr e^{-\beta r^2}\Pi_{j=1,2,...,N}\Big(2\cosh\Big[ \frac{\beta\epsilon}{2}\Big[ 1 + \frac{16{\lambda_j}^2 r^2}{\epsilon^2 N} \Big]^{1/2} \Big]\Big).
    \end{align}
    Defining the variable $y=\frac{r^2}{N}$ allows us to rewrite the above integral as: 
    \begin{align}
      Z(N, T) &= N\int_0^{\infty}dy e^{-\beta N y}\Pi_{j=1,2,...,N}\Big(2\cosh\Big[ \frac{\beta\epsilon}{2}\Big[ 1 + \frac{16{\lambda_j}^2 y}{\epsilon^2} \Big]^{1/2} \Big]\Big)\nonumber\\
      &= N\int_0^{\infty}dy\exp\Big( -\beta N y + \sum_{j=1}^{N}\log\Big[ \Big(2\cosh\Big[ \frac{\beta\epsilon}{2}\Big[ 1 + \frac{16{\lambda_j}^2 y}{\epsilon^2} \Big]^{1/2} \Big]\Big) \Big] \Big).
    \end{align}
    \end{widetext}
    We can write this more compactly as
    \begin{align}
      Z(N,T) &= N\int_0^{\infty}dy \exp\Big( \phi_N(y) \Big)
    \end{align}
   using a shorthand:
    \begin{equation}
      \phi_N(y) =\nonumber\\
      -\beta N y + \sum_{j=1}^{N}\log\Big[ \Big(2\cosh\Big[ \frac{\beta\epsilon}{2}\Big[ 1 + \frac{16{\lambda_j}^2 y}{\epsilon^2} \Big]^{1/2} \Big]\Big) \Big].
    \end{equation}
    for the exponent. We would like to evaluate the above integral
    using the method of steepest descent for which we would like to
    extract the point at which $\phi_N(y)$ is a maximum. To find this,
    we compute the derivative:
    \begin{align}
      \frac{d\phi_N(y)}{dy} &= -\beta N + \frac{4\beta}{\epsilon}\sum_j\frac{\lambda_j^2}{\eta_j}\tanh\Big( \frac{\beta\epsilon\eta_j}{2} \Big), 
    \end{align} 
    where we are using the shorthand notation 
    \begin{align}
      \eta_j &= \Big[ 1 + \frac{16{\lambda_j}^2 y}{\epsilon}^2\Big]^{1/2}.
    \end{align}
    A vanishing derivative, $\frac{d\phi_N(y)}{dy}= 0$, implies: 
    \begin{align}
      0 &= -\beta N + \frac{4\beta}{\epsilon}\sum_j\frac{\lambda_j^2}{\eta_j}\tanh\Big( \frac{\beta\epsilon\eta_j}{2} \Big).
    \end{align}
    Following the intuition from the corresponding calculation for the
    clean Dicke model, we argue that the critical value of the inverse
    temperature must correspond to the case when all the $\eta_j$ take
    their minimum possible value namely unity.  
    Inserting $\eta_j = 1$,
    we have:
    \begin{align}
      0 &= -\beta_c N + \frac{4\beta_c}{\epsilon}\sum_j\lambda_j^2\tanh\Big( \frac{\beta_c\epsilon}{2} \Big)
    \end{align}
    which can be reshaped into
    \begin{align}
      \sum_j\lambda_j^2\tanh\Big( \frac{\beta_c\epsilon}{2} \Big) &= \frac{N\epsilon}{4}.
    \end{align}
    This gives us
    \begin{align}
      \tanh\Big( \frac{\beta_c\epsilon}{2} \Big) &= \frac{\epsilon}{4\frac{\sum\lambda_j^2}{N}}= \frac{\epsilon}{4\langle\lambda^2\rangle}.
    \end{align}
    Substituting for the expressions for $\epsilon$ and $\lambda$, we
    obtain an expression for the transition temperature:
    \begin{align}
      T_c &= \frac{\omega_0}{2\omega}\frac{1}{\tanh^{-1}\Big( \frac{\omega_0\omega}{4\langle g^2\rangle} \Big)}.
      \label{eqn:Tc}
    \end{align} 
    
    To verify this expression for the critical temperature, we
    numerically study the mutual information between two spins, which
    has been shown to be a useful marker for the thermal phase
    transition in the Dicke model~\cite{das2022revisiting,
      das2023phase}.  In Fig.~\ref{fig:MI}, we show the mutual
    information between two spins as a function of the mean coupling
    strength $\mu$ and the temperature $T$, for different standard
    deviations: $\sigma=0.2,\ 0.4,\ 0.5,\ 0.8$ . For $\sigma = 0.2$
    the phase diagram is almost identical to the one for the usual DM
    (see Fig. 4(c) of our earlier work~\cite{das2022revisiting}). For
    $\mu<\frac{1}{2}$
    the system lies in the normal phase, which 
    gives rise here to the black color; for $\mu>\frac{1}{2}$, 
    there is a thermal phase transition from the super-radiant phase 
    (light color) to the normal phase around some critical temperature. 
    In this figure the red dashed 
    line denotes the analytical 
    critical temperature of Eq.~\ref{eqn:Tc}. We can see that it describes 
    the numerical results well. If the standard deviation is increased, it 
    is clear from Fig.~\ref{fig:MI} (b) ($\sigma = 0.4$) and (c) ($\sigma = 
    0.5$) that the thermal phase transition starts 
    at lower mean values $\mu$ than $\mu=g_c$. Finally, for 
    sufficiently wide coupling distributions, with e.g.~$\sigma=0.8$, there 
    is a clear TPT from the SP to the NP even for vanishing mean coupling 
    strength $\mu=0.0$. Hence, 
    we can conclude from Fig.~\ref{fig:MI} that if we introduce disorder with a sufficiently broad distribution into the coupling 
    strength between spins and bosons, 
    there exists a TPT even for vanishing mean coupling
    $\mu=\langle g\rangle=0$.
    
    In Fig.~\ref{fig:MI2}, we show 
    similar data, but using the distribution based on angles, \eref{angle_dist}-\eref{angle_to_coupling}. We again show the mutual information
    between two spins as a function of the mean and the standard
    deviation of the spin-boson coupling strength for fixed temperatures. 
    Here $\theta$ is a random number, drawn from a Gaussian distribution 
    and $g=2\cos\theta$ which we describe in the subsection~\ref{sec_3c}. 
    In the normal phase the mutual information is 
    small, shown by 
    black color. On the other hand, in the super-radiant phase $I_{12}$ 
    is relatively high, which is represented by the other colors. One can 
    notice that when we gradually increase the temperature, the normal 
    phase is also 
    expanding in parameter space. In this figure the white dashed curves 
    represent the critical values of $\sigma$ and $\mu$ for TPT that we derived analytically in \eref{eqn:Tc}, which 
    separate the normal and super-radiant phases quite well. 
    
    \section{Realizing disordered couplings in Dicke model with cold atom or ultracold molecules in a cavity}\label{sec_5}
    While we intentionally study the abstract model \bref{eqn:H_DDM} 
    such that it can apply to a variety of systems, we provide in this
    section some examples for practical realisations.  When
    considering the origin of the Hamiltonian \bref{eqn:H_DDM} through
    light matter coupling for two-level systems in a single mode
    optical cavity, one has, in the dipole
    approximation~\cite{dowling2003quantum}
%
\begin{align}
\label{origin_coupling}
g_i &=\sqrt{\frac{1}{2\hbar\epsilon_0\omega}}\omega_0 u(\mathbf{x}_i) d_{21} \cos{\theta}_i,
\end{align}
where $\epsilon_0$ is the vacuum permeability, $u(\mathbf{x}_i)$ the
cavity mode amplitude at the location $\mathbf{x}_i$ of the $i^{th}$
two level system, and $d_{21} \cos{\theta}_i$ the transition dipole
matrix element between $\ket{2}$ and $\ket{1}$ projected onto the
local cavity field axis, where we have made the dependence on the
angle $\theta_i$ between cavity field at $\mathbf{x}_i$ and transition
dipole axis explicit.

Even for identical atoms or molecules, treated as an approximate
two-level system (TLS), a random position distribution $\mathbf{x}_i$
can now translate into disordered coupling strengths through the position of the TLS relative to the cavity field structure in
$u(\mathbf{x}_i)$ that may contain standing waves, which will cause
disorder in the field strength. 
While this can easily be avoided by trapping all atoms on spatial
scales small compared to the cavity wavelength
\cite{Neuzner_lattice_in_cavity_NaturePhotonics}, one can just as well
generate a range of coupling distributions by weakly trapping the
atoms on the flanks of a standing wave
\cite{Puppe_trapping_in_cavity_PhysRevLett}.

For atomic TLSs without any external fields other than the cavity
field, there would be no additional contribution from the transition
dipole orientation, since we can always choose the quantisation axis
along the local cavity mode electric field direction, such that
$\cos{\theta}_i\rightarrow1$. This is no longer true once an
additional external field perturbs the symmetry, or the particle is
asymmetric, such as most molecules are.

A symmetry breaking field $\mathbf{B}$ could be magnetic, strong
enough to Zeeman-shift undesired magnetic sublevels of the excited
state out of cavity resonance and locally defining the quantisation
axis. If the cavity is penetrated for example by the circular magnetic
field around a current carrying wire, a random 3D distribution of
atomic positions will translate into a random distribution of angles
between quantisation axis and cavity mode electric field, and hence
affect couplings, as sketched in \fref{fig:schematic}.

Another approach to break the symmetry of the two-level system would
be considering ultra-cold molecules~\cite{krems2005molecules,
  carr2009cold} in the optical
cavity~\cite{herrera2016cavity}. Typical hetero-nuclear molecules
possess transition dipole moments with a fixed orientation relative to
the molecular axis \cite{badanko2018communication}.  Molecules
oriented randomly in 3D, such as in the ground state with angular
momentum $J=0$ of the quantum mechanical rotor, will thus exhibit a
distribution of couplings.  Disadvantages of molecules are their
vibrational and rotational degrees of freedom, which are undesired
here. However eliminating or minimising the impact of these is also
required for quantum information and quantum simulation applications
of ultra-cold molecules and aids cooling them, and is thus being
actively pursued. Coupling to both degrees of freedom can be strongly
suppressed, by choosing a molecule with a nearly diagonal
Franck-Condon factor~\cite{hao2019high} between ground and excited
state, and a larger angular momentum in the ground state than the
excited state~\cite{stuhl2008magneto, shuman2010laser}.

Randomly oriented molecules neatly realise the uniform coupling
distribution that we focussed on, since the probability of a given
polar angle $\theta$ is $P(\theta) = \frac{\sin\theta}{2}$ and hence
$P(\cos\theta)$ will be uniform. Refined distributions can then be
tailored by partially orienting molecules along the cavity field axis,
e.g. $P(\theta)={\cal N}e^{-(\theta-\theta_0)^2/\sigma_\theta^2}$,
with $\theta_0$ enforced by an additional external bias field
$\mathbf{E}$ (see \fref{fig:schematic}).
    
The implementations of the disordered Dicke model with cold atoms and
molecules in cavities that we discussed above can then provide
controlled insight, which can be leveraged for understanding the
underlying Hamiltonian in more complex and less controlled cases, such
as when studying superradiance effects in semi-conductor quantum dot
lattices~\cite{Raini_superfluor_quantumdots_Nature}.  In this case
transition dipoles of carriers in quantum dots are also likely
disordered by additional fields and quantum dot geometries, however a
clear distinction of such effects from other disorder and decoherence
sources will be much more difficult.
    
\section{Summary and conclusions}\label{sec_6}
%
In this work, we propose and investigate a disordered single-mode
Dicke model. We specifically focus on two concrete random
distributions of the spin-boson coupling parameters $g_i$: (i) a
uniform distribution, (ii) $g_i\propto\cos\theta_i$, where $\theta_i$
are Gaussian random variables and study the resulting quantum and
thermal phase transitions in the disordered Dicke model. In both cases
we see similar results and hence demonstrate that our results are
robust to changes in the detailed shape of the distribution.
     
We find that the phase transitions depend on both the mean and the
standard deviation of the random coupling strengths. For the QPT, we
find that for mean coupling strengths significantly smaller than their
standard deviation $\sigma$, the latter plays a role similar to the
coupling $g$ in the clean Dicke model. Even for vanishing mean
coupling $\mu=0$, the system thus shows a QPT around $\sigma =
\frac{0.5}{\sqrt{3}}$ for uniformly distributed couplings. When $\mu$
is systematically increased, the critical value of $\sigma$ decreases
and after a certain mean coupling ($=g_c$) the QPT disappears. We
derive approximate expressions for the ground state energy and the
average boson number around the critical line: $\mu + \sqrt{3}\sigma =
\frac{1}{2}$, for the QPT in the $\mu-\sigma$ plane, which provide a
reasonably good and simple approximation of our numerical
results. Exploiting a symmetry of the Dicke model allows us to improve
on the expression.

We also derive an analytical expression for the critical temperature
and numerically verify it with the aid of mutual information between
two spins. It shows that for wide distributions, such that $\sigma$ is
large, there is a phase transition from SP to NP at $\sigma\approx
0.8$ even for vanishing mean coupling strength, $\mu = 0$.
    
The disordered Dicke model should describe quantum dot superlattices
in semiconductor quantum optics (see
e.g.~\cite{Raini_superfluor_quantumdots_Nature}).  Additionally, we
list several methods, by which the disordered Dicke model can be
realized in experiments with ultracold atoms or molecules in a cavity.

\section*{Acknowledgments}
We are grateful to the High Performance Computing (HPC) facility at
IISER Bhopal, where large-scale calculations in this project were
run. P.D. is grateful to IISERB for the PhD fellowship.  A.S
acknowledges financial support from SERB via the grant (File Number:
CRG/2019/003447), and from DST via the DST-INSPIRE Faculty Award
[DST/INSPIRE/04/2014/002461].

\appendix

\section{The uniform distribution}    
\label{app_uniform_calculations}

Consider the scenario where the coupling $g$ is drawn from a
uniform distribution:
\begin{equation}
  P_{u}(g) = \begin{cases}\frac{1}{2\sqrt{3}\sigma} \quad \textrm{if}\quad \mu-\sqrt{3}\sigma < g < \mu+\sqrt{3}\sigma\\
    0 \quad\quad\quad \textrm{otherwise}
  \end{cases}
  \label{app_eqn:uniform}     
\end{equation}
with mean $\mu$ and standard deviation $\sigma$.

To calculate the disorder-averaged ground state energy and average boson number, we have to evaluate: 
\begin{align}
  \overline{E_G} &= \int_{x_1}^{x_2} P_{u}(g) E_G dg,\label{app_eqn:average1}\\
  \overline{\langle a^{\dagger}a\rangle} &= \int_{x_1}^{x_2} P_{u}(g) \langle a^{\dagger}a\rangle dg,  
  \label{app_eqn:average2}
\end{align}
where $E_G$ is given in Eqn.~\ref{eqn:E_gs_dm} and $\langle
a^{\dagger}a\rangle$ is given in Eq.~\ref{eqn:boson_dm}. The lower and
upper limits of the box distribution are: $x_1 = \mu - \sqrt{3}\sigma$
and $x_2 = \mu + \sqrt{3}\sigma$ respectively and we consider $\mu$
and $\sigma$ to be in the range: $[0,1]$. Depending on the relation
between $x_1$ and $x_2$ and $g_c$, there are five cases to be
considered.  After performing the integration outlined above, we
obtain an expression for the disorder averaged ground state energy:
\begin{widetext}
  \begin{align}
    \overline{E_G} &= 
    \begin{cases}
      D_1\left[ \frac{2g_c}{3} - \frac{x_1^3}{3} + \frac{g_c^4}{x_1} \right]              & x_1<-g_c\ \text{and}\ 0<x_2\leq g_c\\
      D_1\left[ \frac{4g_c^3}{3} + \frac{1}{3}\left( x_2^3 - x_1^3 \right) + g_c^4\left( \frac{1}{x_1} - \frac{1}{x_2} \right) \right]              & x_1<-g_c\ \text{and}\ x_2>g_c\\
      - \frac{N\omega_0}{2}                & |x_1|<g_c\ \text{and}\ 0<x_2\leq g_c\\
      D_1\left[ \frac{x_2^3}{3} - \frac{g_c^4}{x_2} + \frac{2g_c}{3} \right]              & |x_1|<g_c\ \text{and}\ x_2>g_c\\
      D_1\left[ \frac{1}{3}\left( x_2^3 - x_1^3 \right) + g_c^4\left( \frac{1}{x_1} - \frac{1}{x_2} \right) \right]              & x_1,\ x_2>g_c\ \text{and}\ \ x_1<x_2
    \end{cases}
    \label{app_eqn:E_gs_disorder_dm} 
  \end{align}
  where $D_1 = - \frac{N}{2\sqrt{3}\sigma}$. 
  For the average boson number the disorder-averaged expression is:
  \begin{align}
    \overline{\langle a^{\dagger}a\rangle} &= 
    \begin{cases}
      D_2\left[ - \frac{4g_c^2}{3} - \frac{x_1^3}{3g_c^2} - \frac{g_c^4}{x_1} \right]              & x_1<-g_c\ \text{and}\ 0<x_2\leq g_c\\
      D_2\left[ - \frac{8g_c^3}{3} + \frac{1}{3}\left( x_2^3 - x_1^3 \right) + g_c^4\left( \frac{1}{x_2} - \frac{1}{x_1} \right) \right]              & x_1<-g_c\ \text{and}\ x_2>g_c\\
      0                & |x_1|<g_c\ \text{and}\ 0<x_2\leq g_c\\
      D_2\left[ \frac{x_2^3}{3} + \frac{g_c^4}{x_2} - \frac{4g_c^3}{3} \right]              & |x_1|<g_c\ \text{and}\ x_2>g_c\\
      D_2\left[ \frac{1}{3}\left( x_2^3 - x_1^3 \right) + g_c^4\left( \frac{1}{x_2} - \frac{1}{x_1} \right) \right]              & x_1,\ x_2>g_c\ \text{and}\ \ x_1<x_2
    \end{cases}
    \label{app_eqn:boson_disorder_dm} 
  \end{align}
\end{widetext} 
where $D_2 = \frac{N}{2\sqrt{3}\sigma\omega^2}$.

For the NP ($|g|\leq g_c$) the third case is applicable and thus:
\begin{align}
  \overline{E_G} &=  -\frac{N\omega_0}{2},\\
  \overline{\langle a^{\dagger}a\rangle} &= 0.
\end{align} 
On the other hand, for the SP ($|g| > g_c$), we consider only the
fourth case: $|x_1|<g_c\ \&\ x_2>g_c$ for the QPT around $g_c$. Thus:
\begin{align}
  \overline{E_G} &= - \frac{N}{2\sqrt{3}\sigma} \left[ \frac{x_2^3}{3} - \frac{g_c^4}{x_2} + \frac{2g_c}{3} \right],\\
  \overline{\langle a^{\dagger}a\rangle} &= \frac{N}{2\sqrt{3}\sigma\omega^2} \left[ \frac{x_2^3}{3} + \frac{g_c^4}{x_2} - \frac{4g_c^3}{3} \right].
\end{align}

\bibliography{ref}                                      
\end{document}